\documentclass[12pt,preprint]{aastex}

\def\lya{Ly$\alpha $}

\def\ergcm2s{\ifmmode {\rm\,erg\,cm^{-2}\,s^{-1}}\else
                ${\rm\,ergs\,cm^{-2}\,s^{-1}}$\fi}

\def\Mpc{\hbox{Mpc}}

\begin{document}

\title{An Overdensity of Lyman-$\alpha$ Emitters at
Redshift z $\approx$ 5.7 near the Hubble Ultra Deep Field}
\author{J. X. Wang\altaffilmark{1,3}, S. Malhotra\altaffilmark{2},
J. E. Rhoads\altaffilmark{2}}
\begin{abstract}

We have identified an obvious and strong large scale structure at
redshift $z\approx 5.75$ in a wide (31$\arcmin\times$33$\arcmin$)
field, narrowband survey of the Chandra Deep Field South region.  This
structure is traced by 17 candidate \lya\ emitters, among which 12 are
found in an 823nm filter (corresponding to \lya\ at $z=5.77 \pm 0.03$)
and 5 in an 815nm image ($z=5.70 \pm 0.03$).  The \lya\ emitters in
both redshift bins are concentrated in one quadrant of the field.  The
Hubble Ultra Deep Field, Chandra Deep Field South, and GOODS-South
fields all lie near the edge of this overdensity region.  Our results
are consistent with reports of an overdensity in the UDF region at
$z\approx 5.9$. This structure is the highest redshift overdensity
found so far.

\end{abstract}


\altaffiltext{1}{Center for Astrophysics, University of Science and Technology of China, Hefei, Anhui 230026, P. R. China; jxw@ustc.edu.cn.}
\altaffiltext{2}{Space Telescope Science Institute, 3700 San Martin Drive, Baltimore, MD 21218; san@stsci.edu, rhoads@stsci.edu.} 
\altaffiltext{3}{Department of Physics and Astronomy, Johns Hopkins University, 3400 N. Charles Street, Baltimore, MD 21218.}
\section {Introduction}

The study of the large-scale clustering of high redshift galaxies is
essential to understand the formation and evolution of galaxies. To
date, many possible detection of large-scale clustering of high
redshift galaxies have been reported (e.g., Steidel et al. 1998, 2000;
Venemans et al. 2002, 2004; Miley et al. 2004; Palunas et al. 2004; Ouchi et
al. 2001, 2003, 2004; Foucaud et al. 2003; and references therein).

The \lya\ emission line searches have been very effective at finding
high redshift galaxies, especially at $z > 5$ where most of the
spectroscopically confirmed galaxies show \lya\ emission.  These
include our Large Area Lyman Alpha (LALA) survey (e.g., Rhoads et
al. 2000) and other recent searches (Cowie \& Hu
1998; Hu et al 1998, 2002, 2004; Kudritzki et al 2000; Fynbo, Moller,
\& Thomsen 2001; Pentericci et al. 2000; Stiavelli et al. 2001; Ouchi
et al. 2003; Fujita et al. 2003; Shimasaku et al. 2003; Kodaira et
al. 2003, Ajiki et al. 2004; Kurk et al. 2004; Tran et al. 2004, and 
references therein).  Large-scale clustering of high
redshift \lya\ emitters has also been reported in some
of these surveys (Venemans et al. 2002, 2004; 
Ouchi et al. 2001, 2003, 2004).

In this letter we report the detection of a large-scale structure of
\lya\ emitters at $z \sim$ 5.7 in $Chandra$ Deep Field South (CDF-S).
The candidate \lya\ emitters at $z \sim$ 5.7 were selected using deep
narrowband images, following the selection criteria developed
in the Large Area Lyman Alpha survey (Rhoads et al. 2000,
2001, 2003, 2004).
Throughout this paper, we assume a cosmology with
$H_0$ = 71 km s$^{-1}$ Mpc$^{-1}$, $\Omega_M$ = 0.27 and
$\Omega_\Lambda$ = 0.73 (cf. Spergel et al. 2003).

\section{Deep Narrow Band Imaging}

We imaged the area around Chandra Deep Field South using two
narrowband filters with central wavelength $\lambda_c$ of 815 and 823
nm. The full width at half maximum (FWHM) transmissions of the two
filters is 7.5 nm, with peak throughputs of $\sim 90\%$. 
These were new filters of high quality.  In particular,
their throughput does not vary substantially across the field
of view (unlike the similar but older filters used by
Rhoads \& Malhotra 2001).
The deep
narrowband images were obtained using the Mosaic II CCD imager at the
CTIO 4m V. M. Blanco telescope, on 2003 Dec. 22--30 (UT).  Chip 3 of
the 8 SITe 2Kx4K CCDs was turned off during the observations due to
high amplifier noise.  Imaging data reduction followed the methods
described in Rhoads et al. (2000, 2004).  To summarize, in addition to
the standard CCD data reduction steps (overscan subtraction, bias
frame subtraction, and flat-fielding), we also removed electronic
crosstalk between Mosaic chip pairs sharing readout electronics, and
the residual, large-scale imperfections in the sky flatness using a
smoothed supersky flat derived from the science data.  Cosmic rays
were rejected in each exposure using the algorithm of Rhoads
(2000). Astrometry of USNO-A2 catalog stars was used to adjust the
world coordinate systems of individual frames.  Satellite trails were
flagged manually for exclusion from the final stacks.  Weights for
image stacking were determined using the ``ATTWEIGHT'' algorithm
(Fischer \& Kochanski 1994), and the task ``mscstack'' in the MSCRED
package (Valdes 1998) was used to stack the individual exposures.

The total integrated exposure time of the final stacked images are
34.7 ks for the 815 nm filter and 36.0 ks for the 823 nm filter. The
final stacked images have seeing of 0.9\arcsec, while the seeing of
823 nm band image is slightly better (0.02\arcsec) than that of 815 nm 
band image.
We also used deep $B,V,I$
broad band images of ESO Imaging Survey (EIS) in the CDF-S, obtained
using the Wide Field Imager (WFI) of the ESO-MPG 2.2 meter telescope
at La Silla. The data were download from ESO archive, and registered to the
narrow band images for candidate selection. The overlap between the narrowband
stacks and the EIS images was $31\times 33$ arcminutes in size with
a gap of $10.5\times 13$ arcminutes due to the missing chip.
The total solid angle searched for \lya\ galaxies
was thus 887 square arcminutes.  The corresponding volume over the
full redshift range, $5.67 \le z \le 5.80$, is $3.06\times 10^5 
\Mpc^3$.
A summary of the images is given in Table 1.

\section{Candidate z= 5.7 \lya\ Emitters}

Source detections were performed using SExtractor version 2.2.2
(Bertin \& Arnouts 1996) on the narrowband images, and 
measured their colors using SExtractor's two-image mode
to measure the EIS $B,V,I$ band
photometry for the narrowband detected sources. All fluxes are
measured in 2.43\arcsec\ (9 pixel) diameter aperture.

Candidate $z \sim 5.7$ \lya\ emitters were selected following the same
criteria that have proven highly successful in the LALA fields (Rhoads \& Malhotra
2001; Rhoads et al. 2003, 2004). These are: 1) Narrowband detection
$>$ 5$\sigma$; 2) A narrowband excess of $>$ 0.75 magnitude, so that
$>$ 50\% of the narrowband flux comes from an emission line; 3)
Significance of the narrowband excess $>$ 4$\sigma$; and 4) $<$
2$\sigma$ detection in filters blueward of the expected Lyman break
location ($B,V$ band in this letter). The success rate of the criteria
has been spectroscopically confirmed to be $\ge$ 70\% 
(Rhoads et al. 2003, Dawson et al. 2004, Dawson et al. 2005 in prep).

\begin{figure}
\plottwo {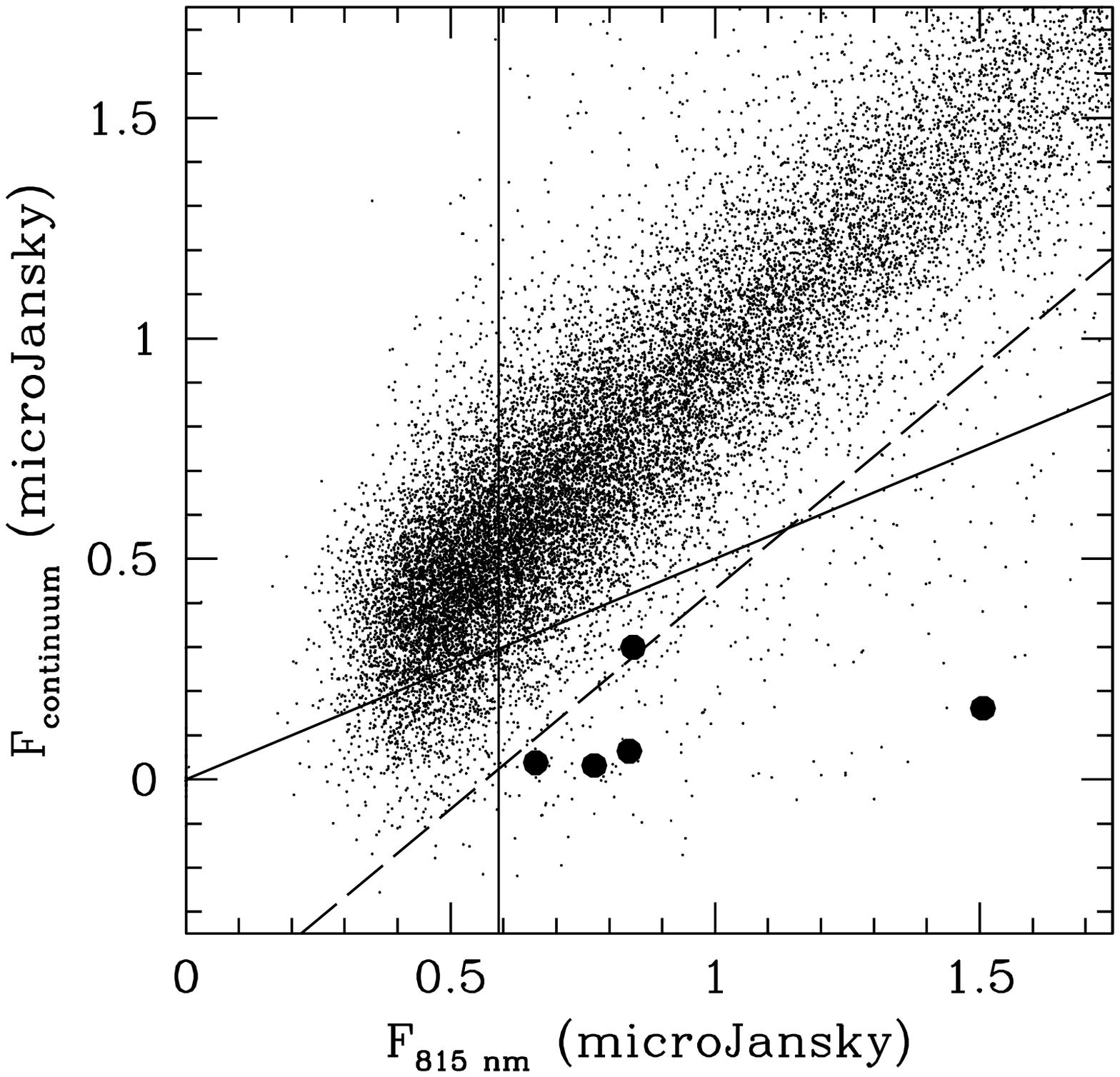}{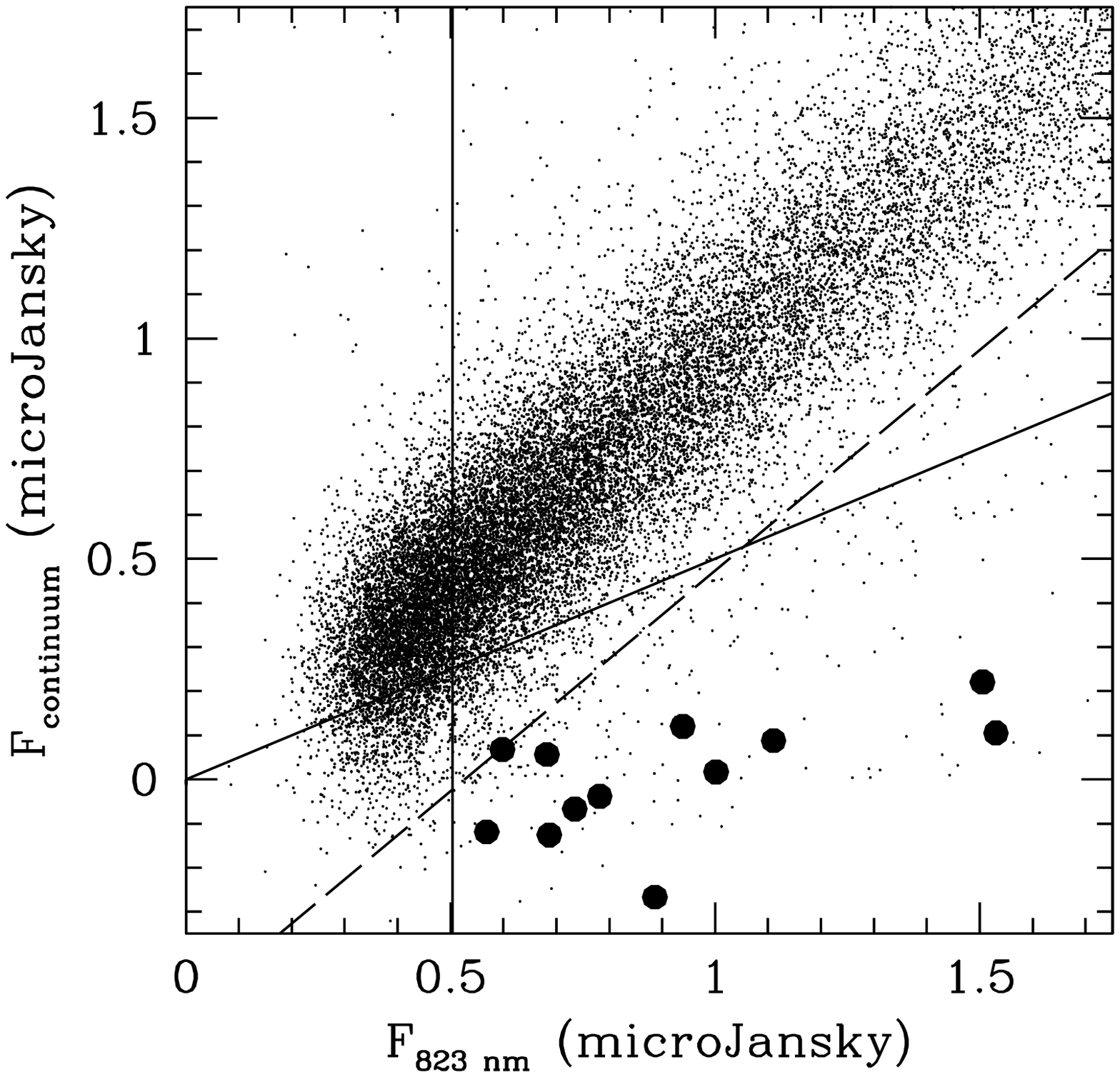}
\caption{Underlying continuum fluxes vs narrow band fluxes for SExtractor
detected sources in the two narrow band images. 
Our candidate $z \sim 5.7$ \lya\ emitters (5 in 815 nm band, and 12 in 823 nm
band) are marked as big dots. Lines show our selection criteria: 
the vertical line stands for 5 $\sigma$ detection in narrow band; the solid 
inclined line represents a narrow excess of 0.75 magnitude ($Flux_{narrow}$/$Flux_{continuum}$ = 2)
and the dashed line corresponds to the requirement of 4$\sigma$ significance
of the narrowband excess which is $Flux_{narrow}$ - $Flux_{continuum}$ = 4$\times\sqrt{(Err_{narrow}^2 + Err_{continuum}^2)}$.
Note there is one candidate in the left panel which does not pass the
three threshold lines in the figure. This is because that the lines are
plotted using the mean uncertainties of continuum and narrow band fluxes, while
these candidates are selected based on their individual flux uncertainties
output by SExtractor, which can be 10-20\% smaller/larger than the mean 
uncertainties.
The negative underlying continuum fluxes in the figure are just for 
display purpose. While selecting candidates, these negative fluxes were
promoted to zero to make our selection conservative.
}
\label{select}
\end{figure}

In order to make full use of our narrow band images, which are deeper
than the $I$ band image, we ran our selection
using the weighted mean flux from $I$ and 815 nm band as the underlying 
continuum for 823 nm selection, and the weighted mean
flux from $I$ and 823 nm band as the continuum for 815 nm selection.
A total of 17 candidates are selected, 5 in the 815 nm band, and 12 in
the 823 nm band. In Fig. \ref{select} we plot the underlying continuum
and narrow band fluxes for all detected sources and candidate $z \sim 5.7$ 
\lya\ emitters selected.
Our selection criteria are also plotted in
the figure: the vertical line stands for 5 $\sigma$ detection in narrow
band; the solid inclined line represents a narrow excess of 0.75 magnitude
and the dashed line corresponds to the requirement of 4$\sigma$ significance 
of the narrowband excess.

\section {Large scale structure}
\begin{figure}
\plotone{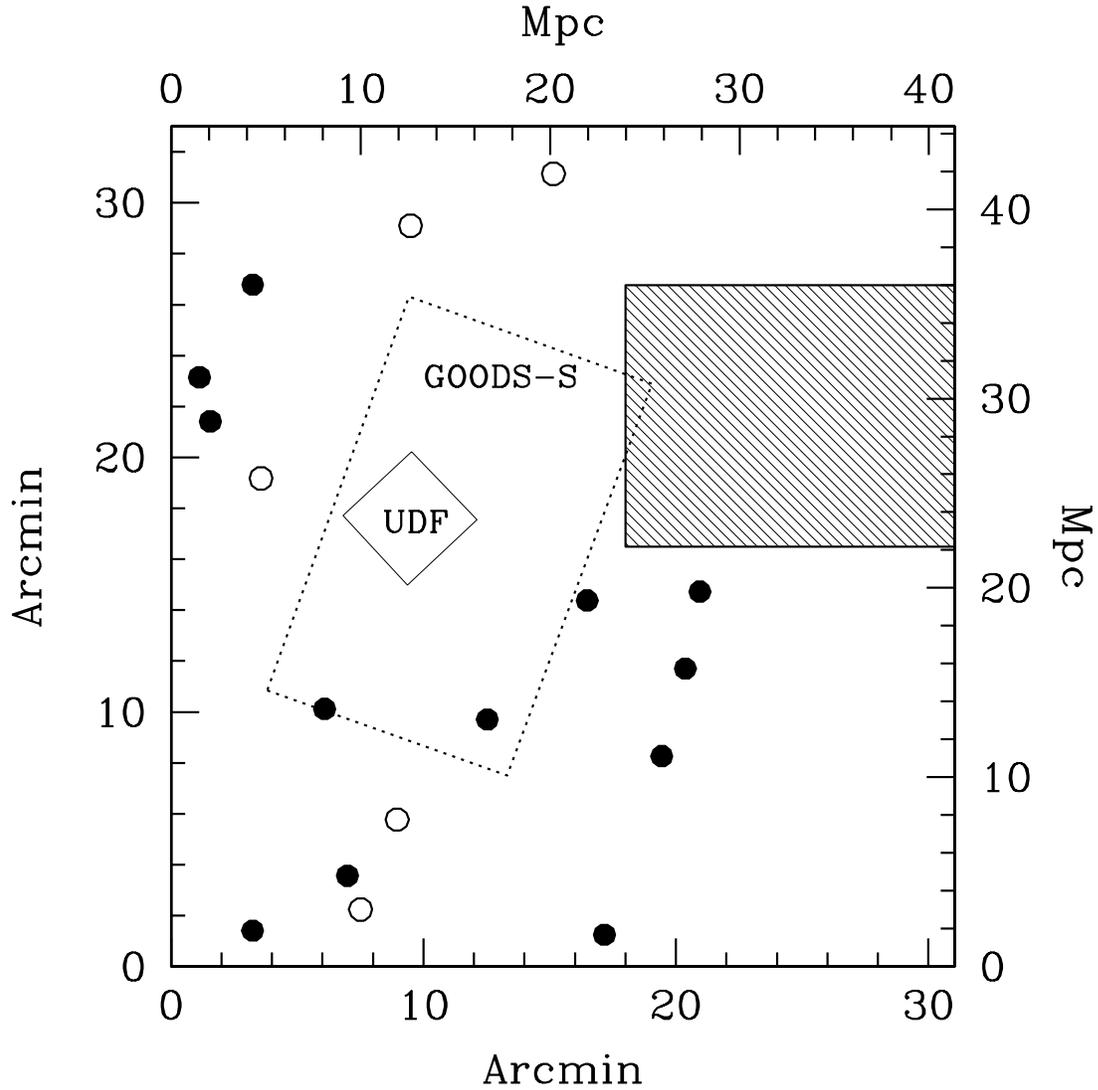}
\caption{
Sky distribution of the 17 photometrically selected $z \sim 5.7$ \lya\ emitters
in CDF-S. The shaded region in the plot is due to the
Chip 3 of the Mosaic II camera which was turned off during the observations.
North is up and east is to the left.
Solid and open circles  are candidates selected from 823 nm and 815 nm band 
respectively. 
See text for details.
}
\label{sky}
\end{figure}

Among the 17 candidate $z \sim 5.7$ \lya\ emitters we detected in CDF-S,
12 of them are selected in the 823 nm band image, 5 of them in the
815 nm band image. 
In Fig. \ref{sky} we plot the sky distribution of the 17
candidate $z \sim 5.7$ \lya\ emitters in CDF-S. An obvious structure 
of the distribution can be seen in the figure: almost all the 17 sources 
are located in the lower left half of the field. 

Before studying the clustering properties of
the 17 \lya\ emitters, we consider whether the structure 
could be an observational artifact.
Such a spurious ``structure'' in the sky distribution
of the selected \lya\ emitters could be due to two reasons:
It might be due to a difference in narrow-band sensitivity
across the field.
It might also due to the sky distribution of very bright
sources in the field, which could affect our source detection
and candidate selection.
 
\begin{figure}
\plotone {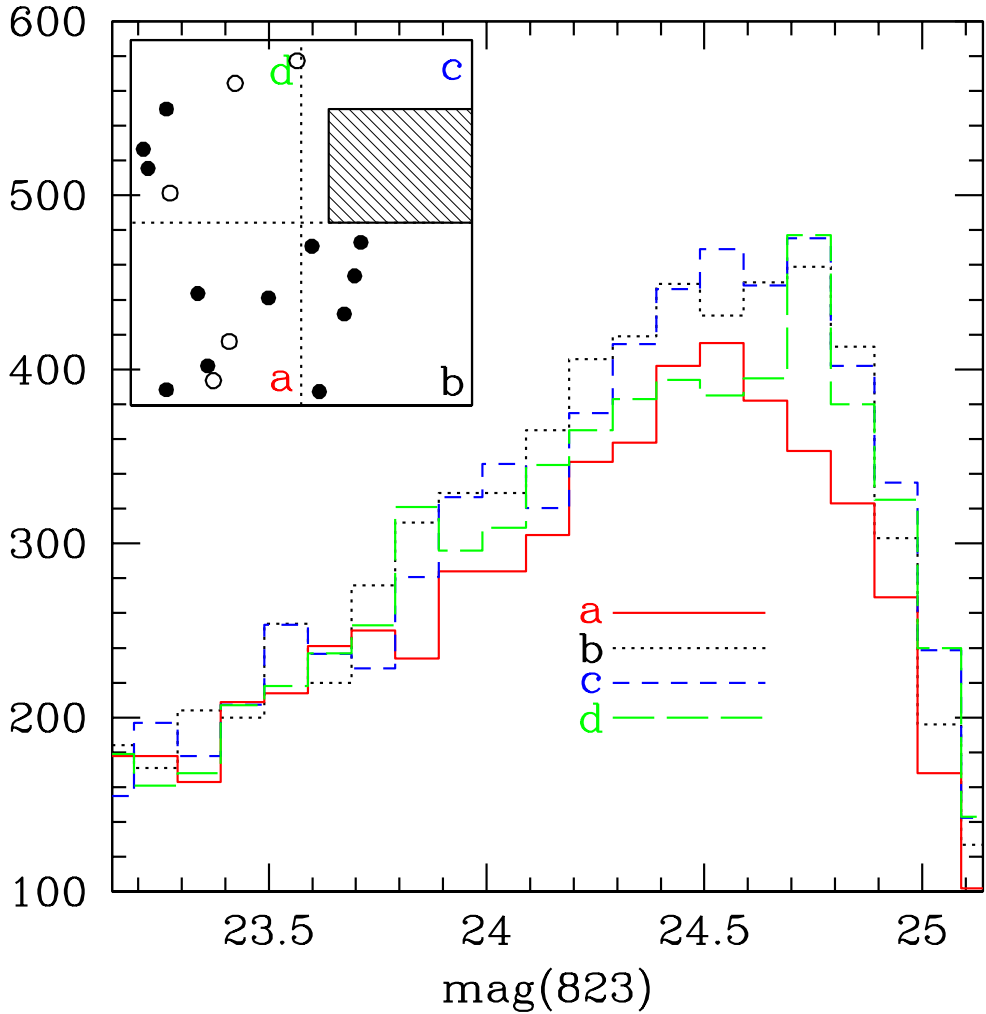}
\caption{
The source number counts in the 823 nm band image, which was
divided into 4 quarters. The numbers of detected sources in
each quarter agrees well with each other, showing that the clustering of 
our $z = 5.7$ candidates is not due to the different detection efficiency
over the field.  (The counts in quadrant~c are adjusted for the reduction
in solid angle due to the missing chip.)}
\label{counts}
\end{figure}

To check these  possibilities, we first divide the 823 nm band image into
four quarters and plot the source counts for the four parts in
Fig.~\ref{counts}. We can see that the source counts in four regions are
consistent within the Poisson counting uncertainties.  
In region~a where we identified the most
\lya\ candidates, the overall source density is among the lowest. 
This indicates that the detection efficiency in region~a is not
higher than in other 3 regions. We found a similar result 
in the 815 nm band image.

Second, we added artificial point sources with narrowband magnitude of 24.3 
(the mean value of the 12 \lya\ emitters in the 823 nm band image)
at random locations in the narrowband image.
We ran exactly the same procedures of source detection and candidate selection
on the derived narrowband images. The recovered artificial sources passing 
our criteria show no evidence of strong clustering.
Inserting artificial sources with different magnitudes (the highest and the
lowest magnitudes of the 12 sources) yields the same results.
Likewise, no sign of clustering is found among artificial
sources inserted and recovered from the 815 nm band image.
A two-dimensional Kolmogorov-Smirnoff
test shows that the sky distribution of the 12 \lya\ emitters at
823 nm is different from the corresponding simulation at
the 99.4\% confidence level.
Including the 5 \lya\ emitters from the 815 nm band would slightly increase the
significance to 99.6\%.
The sky distribution of the 5 \lya\ emitters in 815 nm band alone
is different from the simulated sources at the level of 98\%.
We checked the limiting magnitudes of the 823 nm, $I$, $B$ and $V$ band 
images over the whole field, and found no inhomogeneity.
The photometric zero points of the narrow and broad band images
are also found to be accurate within 0.1 mag over the whole field.
We point out that there is no evidence of significant spatial variations in
bandpass of the narrow band filters either.
All the above checks indicate the clustering of the \lya\ candidates
is not artificial.

\begin{figure}
\plotone {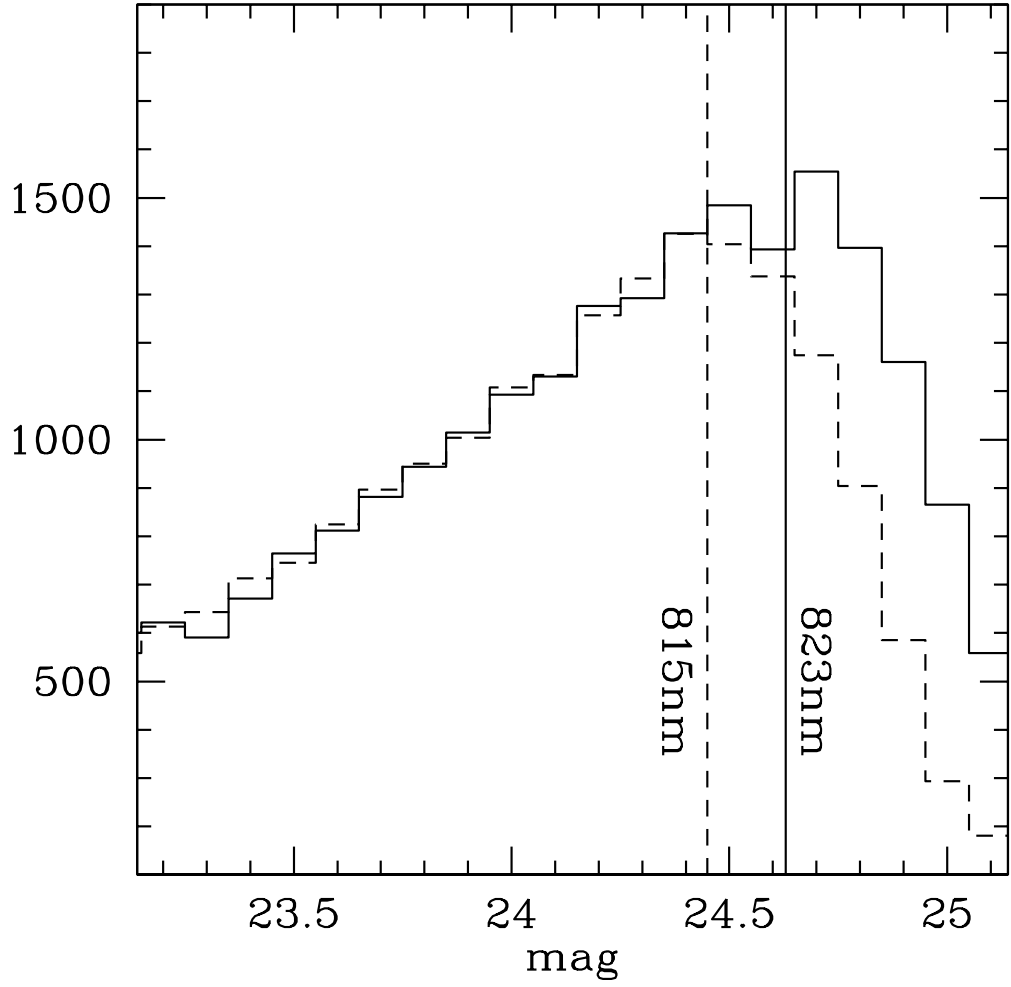}
\caption{Comparing the source number counts in the 815 nm
and the 823 nm band images. The vertical lines stand for
the 5 $\sigma$ detection in the narrow bands as one of
our selection criteria.
}
\label{compare}
\end{figure}

We see clear evidence of large scale clustering of the
12 \lya\ emitters in the 823 nm band. The same trend is
also seen in the 815 nm band, but with lower confidence
level because the number of \lya\ emitters in 815 nm band
is only one third of that in 823 nm band.
In Fig. \ref{compare} we compare the source counts from the
815 nm and 823 nm band images. We can see that the
detection efficiencies are consistent with each other above
the 5 $\sigma$ detection limit, which is 24.45 mag for the
815 nm band image, and $\sim 0.2$ mag deeper for the 823 nm band.
Note there are 2 \lya\ emitters in 823 nm band with narrow band 
magnitudes fainter than 24.45. Removing the 2 sources, the space
density of \lya\ emitters in 823 nm band is still twice 
of that in 815 nm band.
The corresponding limiting Ly$\alpha$ fluxes
are 2.00 \& 1.66 $\times$ 10$^{-17} ergs/cm^{2}/s$ for 
815 nm and 823 nm bands respectively. 
Based the luminosity function
of $z \sim 5.7$ \lya\ emitters given by Malhotra \& Rhoads (2004),
we predict the numbers of \lya\ emitters in the two narrow band
images to be 4.5 \& 7.5. This corresponds to an overdensity factor
of 1.6 for 823 nm band candidates.
In the densest quadrant of the survey, the overdensity factor
reaches 3-4.

\section {Conclusions}
The structure we have found in the \lya\ galaxy distribution at
$z\approx 5.8$ is the most distant large scale feature yet reported in
the spatial distribution of galaxies.  The space density of these
galaxies varies by a factor of two between our two narrowband filters,
and the projected surface densities span a substantially larger range
(up to $4 \times$ the expected mean).

The Hubble Ultra-Deep Field (HUDF; Beckwith et al 2005 in prep) lies near the
edge of this large scale galaxy sheet, and the effects of this
overdensity are also apparent there.  An
overdensity at $z\approx 5.9$ near the HUDF was first suggested by
Stanway et al (2004) on the basis of 3 redshifts in that
region.  This has now been confirmed in the redshift distribution of
23 $i$-dropout Lyman break galaxies in this field (Malhotra et al
2005), as measured using slitless spectroscopy from the GRAPES project
(Pirzkal et al 2004).  The redshift extent of the feature within
the UDF is larger than the coverage of our narrowband
filters, suggesting that the  structure is more complex than
a single sheet.

\acknowledgements 
We would like to thank Bahram Mobasher for providing the calibrated
photometric zero points for the EIS images.
The work of JW was supported by the CAS "Bai Ren" project in USTC
and the CXC grant GO3-4148X.

\clearpage
\begin{deluxetable}{cccc}
\tablecaption{Optical images}
\tablecolumns{4}
\tablewidth{0pt}
\tablehead
  {
  \colhead{Band} & \colhead{Telescope} & \colhead{Exposure Time (ks)} & \colhead{m$_{AB}$(lim)$^a$}
  }
                                                                                
\startdata
$B$  & ESO 2.2m & 24.5  & 27.13 \\
$V$  & ESO 2.2m & 9.2  & 25.97 \\
$I$  & ESO 2.2m & 26.9 &  24.62 \\
815nm & CTIO 4m & 34.7  & 25.03 \\
823nm & CTIO 4m & 36.0  & 25.14  
\enddata
\tablenotetext{a}{
~Limiting magnitudes (AB) for a 3 $\sigma$ detection on a
2.0\arcsec\ diameter aperture.}
\end{deluxetable}
                                                                                
\newpage

\newpage

\end{document}